\documentstyle[11pt,paspconf]{article}
\include{psfig}

\begin{document}

\title{Spectra of Cluster Galaxies at $z\sim 0.4$}
\author{B.M. Poggianti}
\affil{Institute of Astronomy and Royal Greenwich Observatory, Madingley Road,
Cambridge CB3 0EZ, UK}

%\author{A. Dressler}
%\affil{Carnegie Observatories, 813 Santa Barbara Street, Pasadena, CA 91101, 
%USA}

\begin{abstract}
The ``MORPHS'' group has obtained an extensive spectroscopic dataset
of galaxies in the fields of 10 rich clusters at $0.37<z<0.56$;
the stellar population properties of about 700 objects 
have been determined from a detailed analysis of
spectral line strenghts and colours, up to a galactic absolute magnitude
$M_V=-20.3$ mag.
Morphological informations for a consistent fraction of our sample
have been previously derived from deep images of the central regions of 
these clusters taken with the WFPC-2 on board the HST. 
Spectral and morphological features can be compared in order to study 
the star formation histories of the different galactic types in clusters;
here we show some preliminary results of this comparison.

\end{abstract}


\section{Introduction}

Galaxies in rich clusters at low redshifts have stellar and structural
properties significantly different from objects in the nearby field.
Although it is well established that galaxy evolution \sl depends \rm
on the environment, it is still not clear which/how/when/how-much
environmental effects determine or influence the main galactic properties.
In clusters an unexpected strong evolution from intermediate
redshifts to $z=0$ has been detected regarding the star formation activity
(from a large number of studies beginning with those of Butcher \& Oemler)
and, more recently, the relative fractions of the various
morphological types.

The ``MORPHS'' group -- Amy Barger, Harvey Butcher, Warrick Couch, 
Alan Dressler, Richard Ellis, Gus Oemler, Ray Sharples, Ian Smail and 
myself --
has used images from the Hubble Space Telescope Wide Field and Planetary Camera
2 to study the evolution of galaxies in distant clusters. 
The catalog with positions, photometry and
Hubble types is presented in Smail et al. (1997b), while
the morphological evolution and the morphology-density relation
are discussed in Dressler et al. (1997). Constraints on the formation 
epoch of the early type galaxy populations can be placed from
their colour homogeneity (Ellis et al. 1997) and their structural 
parameters (Barger et al.
1997). An analysis of gravitational lensing by the clusters is given in 
Smail et al. (1997a).

Here we report on some first results from a large optical spectroscopic
survey of 10 clusters (see Table), which should be cited
as preliminary results of the MORPH's collaboration.  
The spectral catalog and some basic properties of the sample
will be described in Dressler et al. (1998);
the analysis of the dataset and the interpretation in terms 
of the star formation history will be given in Poggianti et al. (1998).

\section{Analysis}

Redshifts and equivalent widths of the main lines have been measured 
for the $\sim 500$ new spectra, as well
as for the sample of Dressler \& Gunn (1992, DG92).
The spectra have been classified into 6 main spectral classes
according to their emission and absorption line properties and 
have been compared with 
both our spectrophotometric model (Barbaro \& Poggianti 1997)
and a local sample of field galaxies (Kennicutt 1992).
In particular we identified the cases with a recent 
(post--starburst) or current (starburst) strong episode of star formation.

For a subsample of $\sim 300$ galaxies, the \sl spectral type 
\rm can be compared with the \sl morphological type \rm determined 
with the HST: a major goal is to investigate the star formation histories of 
galaxies as a function of their Hubble type and 
explore the connection between the strong \sl morphological \rm
evolution observed (Dressler et al. 1997) and the stellar population
content.
\begin{table}
\caption{Numbers are not definitive and include also 
the DG92 data. The spectra were
taken with the Hale Telescope at Palomar, the William Herschel
Telescope at La Palma and the New Technology Telescope at La Silla; the 
typical spectral range is 3500--8000/10000 \AA}
\begin{center}\scriptsize
\begin{tabular}{lcc}
\tableline
%\hline
\noalign{\smallskip}
Cluster &   $z$   & N spectra \\
\noalign{\smallskip}
%\tableline
        &     & total(cl. members) \\ 
\noalign{\smallskip}
%\hline\noalign{\smallskip}
\tableline
\noalign{\smallskip}
A370      & 0.37  & 61(41) \\
Cl1447+23 & 0.37  & 29(21)  \\
Cl0024+16 & 0.39  & 156(106) \\
Cl0939+47 & 0.41  & 164(71) \\
Cl0303+17 & 0.42  & 94(50) \\
3C295     & 0.46  & 35(25) \\
Cl0412$-$65 & 0.51  & 25(11) \\
Cl1601+42 & 0.54  & 106(56) \\
Cl0016+16 & 0.55  & 42(29) \\
Cl0054$-$27 & 0.56  & 27(13) \\
%\noalign{\smallskip}
%\hline
\tableline
\end{tabular}
\end{center}
%\label{The sample}
\end{table}

\begin{figure}
\centerline{
\psfig{figure=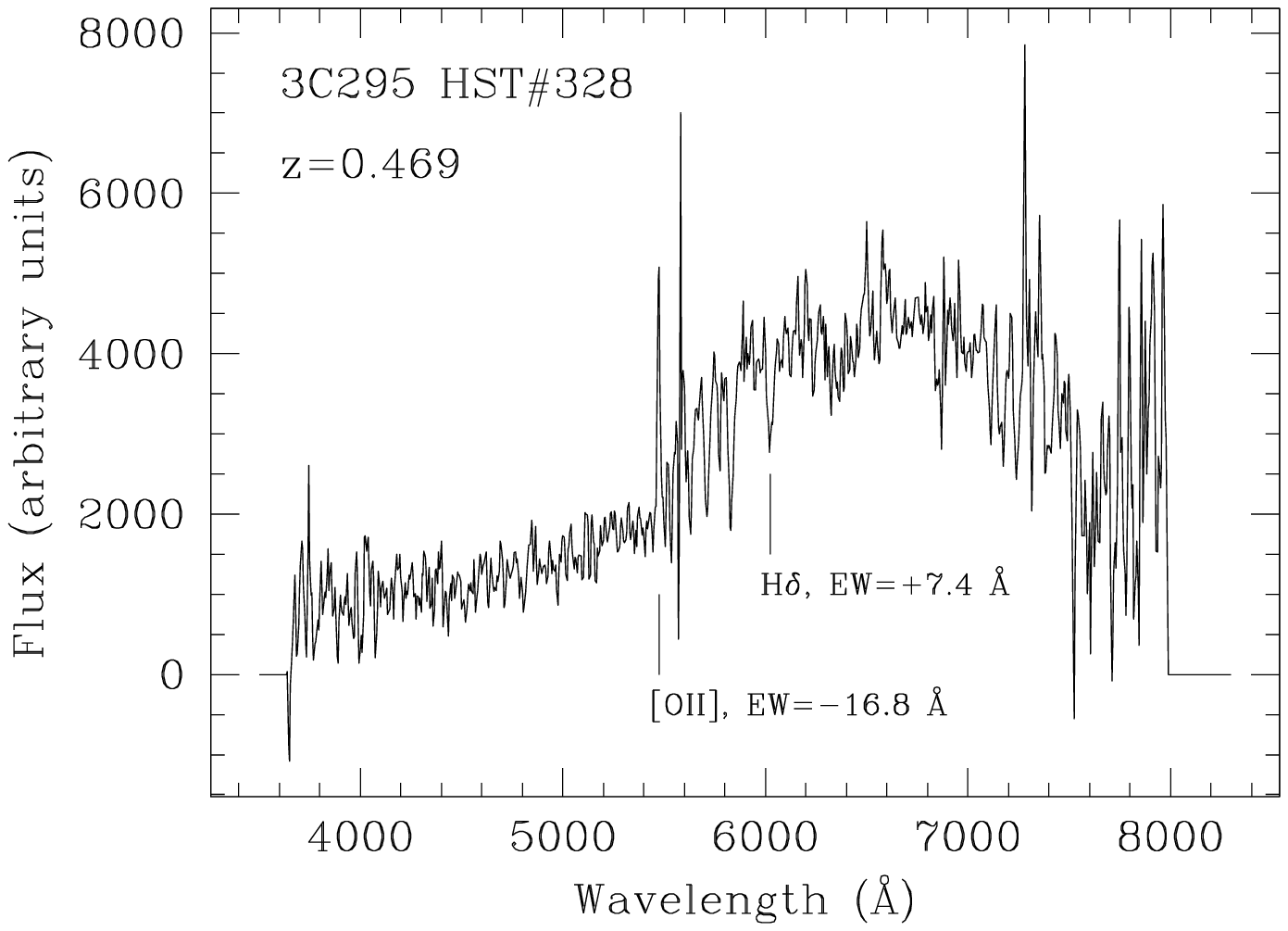,width=8.5cm}
\psfig{figure=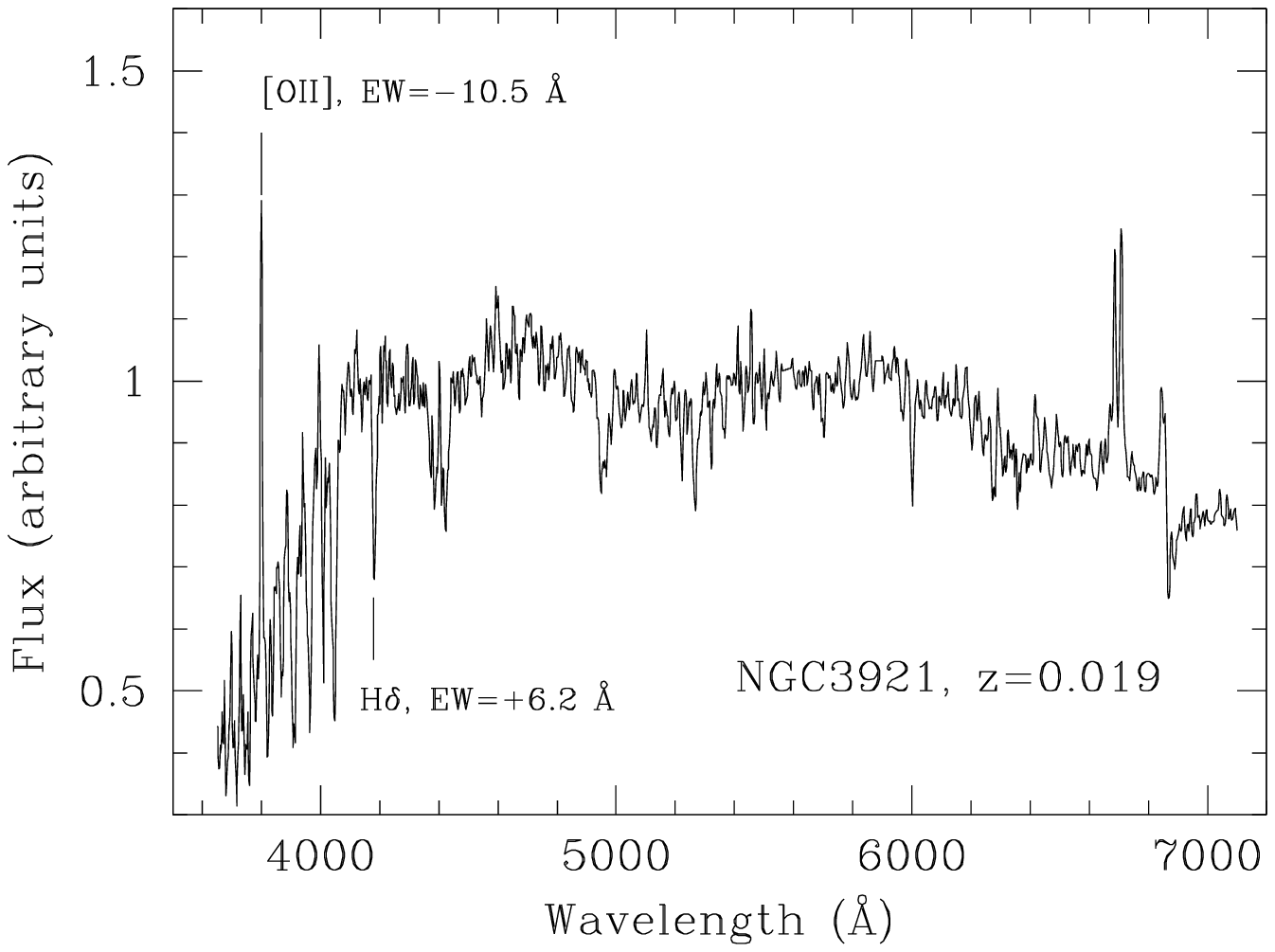,width=8.5cm}}
%\centerline{\psfig{figure=fig1b.ps,width=8cm}}
\caption{The spectrum of the galaxy HST\#328 in 3C295 
is compared with the spectrum of the merging galaxy NGC3921
(Liu \& Kennicutt 1995). They both show a moderate [OII]$\lambda$3727 line 
and a strong $\rm H\delta$ line in absorption.}
\end{figure}

\section{Results}

The main preliminary results for cluster members are the following:

$\rhd$ As in previous spectroscopic surveys of clusters 
at intermediate redshifts (Couch \& Sharples 1987, DG92), 
a large population of post-starburst objects 
is found: about a third of the sample show evidence of a recent starburst which
ended during the last 2 Gyr.

$\rhd$ About 5\% of the spectra have very strong emission lines
indicative of a current starburst. They mostly belong to the lowest 
luminosity end of the sample. If no further star formation is assumed 
after this burst, according to our calculations they will 
fade significantly by $z=0$.

$\rhd$ A large number of spectra (10\%) exhibit exceptionally strong 
higher order Balmer lines in absorption \sl and \rm
the [OII]$\lambda$3727 line in emission. On the basis of our model
we interpret them as objects 
which have had a recent intense starburst
and present some current (residual) star formation activity at 
a much lower level.
There is a number of low redshift examples of similar spectra, such
as those of some merging galaxies (Fig. 1).

\begin{figure}
\centerline{\psfig{figure=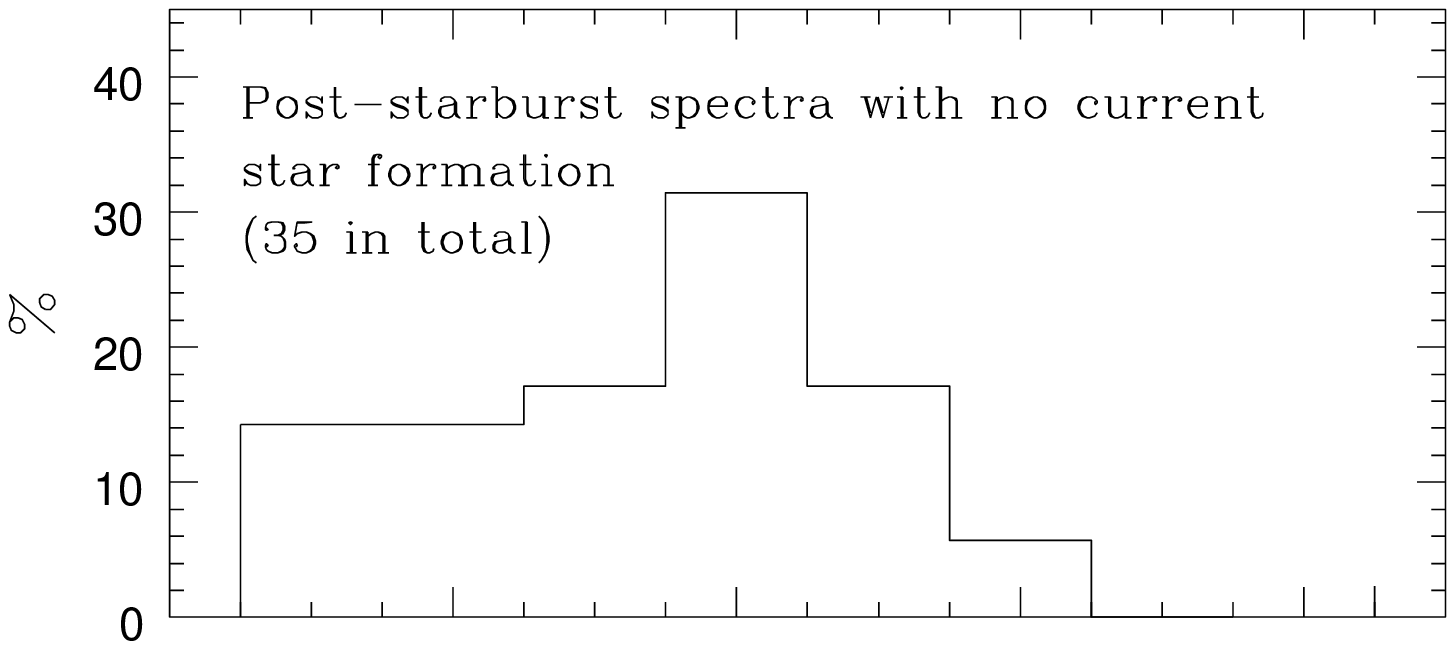,width=10cm}}
\centerline{\psfig{figure=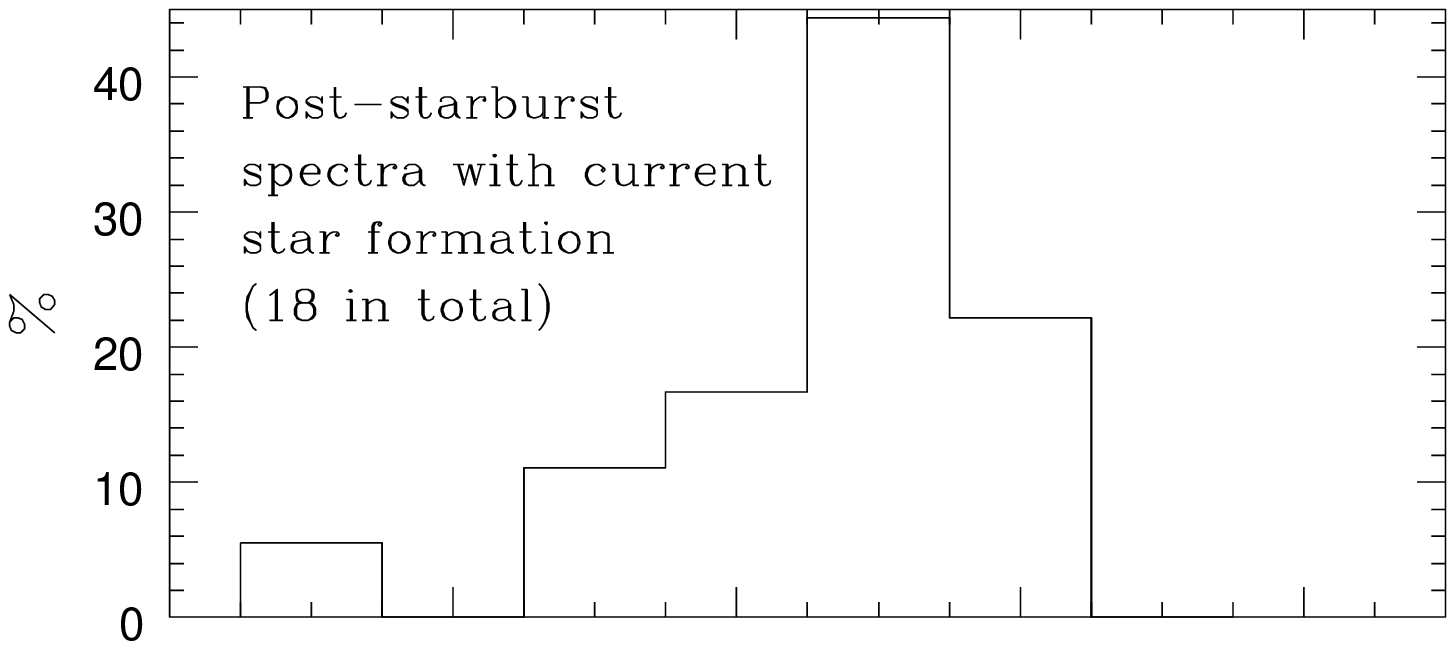,width=10cm}}
\centerline{\psfig{figure=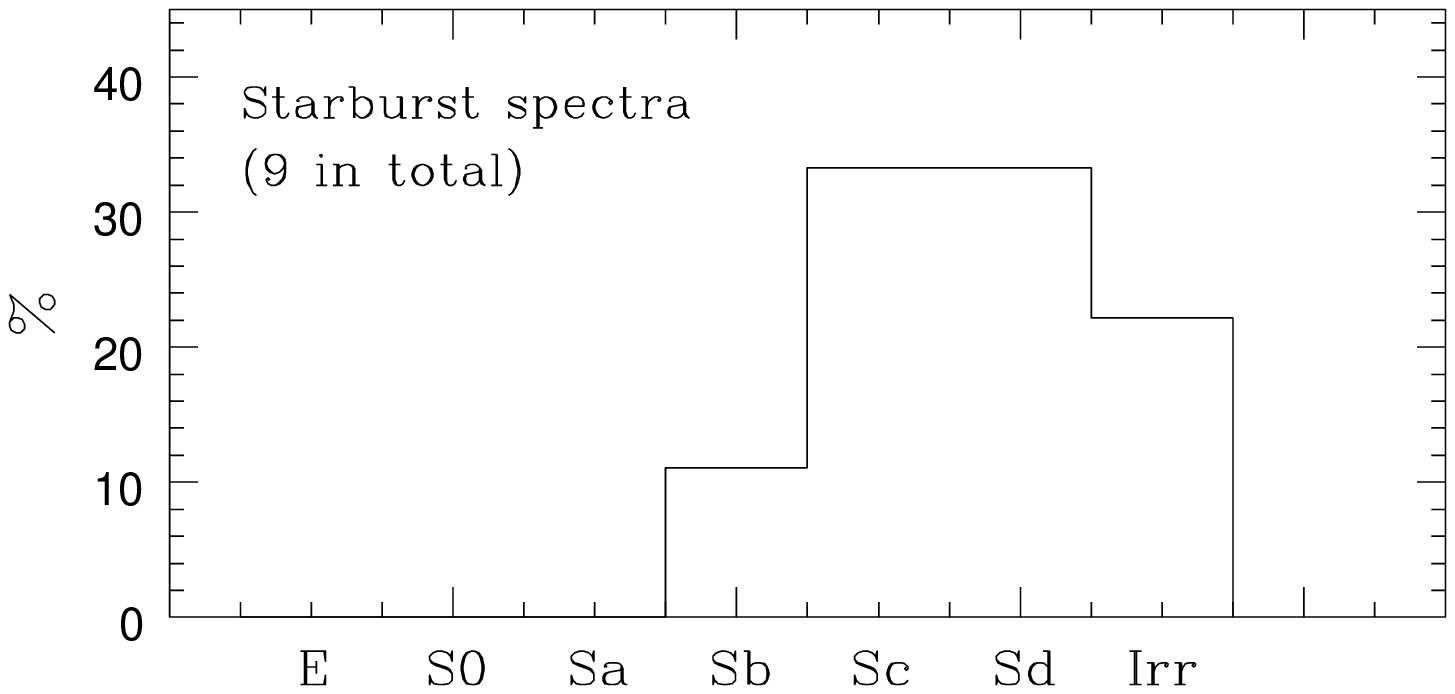,width=10cm}}
\caption{Morphological type distributions for objects
of 3 different spectral classes.}
\end{figure}

$\rhd$ The bulk of the early type galaxies (ellipticals and the few S0's
found in these clusters) have passive spectra, with no obvious signs of
current or recent star
formation. However a non-negligible fraction of them have a post-starburst
spectrum, with strong Balmer lines in absorption and no trace of 
ongoing star formation.

$\rhd$ Of the many spirals, the majority have spectra which are either
``too active'' or ``too passive'' as compared to low--redshift field spirals.
The entire population of starburst galaxies and most of the
post-starburst population is composed of spirals (Fig.2). 
Another significant fraction of the late type galaxies present a level
of star formation activity which is too low for their Hubble type
compared to local field spirals.

$\rhd$ 
Morphologically disturbed objects -- asymmetric or distorted --
are numerous in these clusters (Smail et al. 1997) but
except for a few clear cut examples of mergers, 
the interpretation of the cause of the disturbance remains
subjective.
Most of the starburst spectra and a large fraction of the 
post-starburst spectra belong to galaxies which
appear morphologically strongly disturbed.

\acknowledgments

This work was supported in part by the Formation and Evolution
of Galaxies network set up by the European Commission under contract
ERB FMRX-CT96-086 of its TMR programme.

\end{document}